\documentclass[aip, apl, preprint, amsmath,amssymb]{revtex4-1}

\usepackage{graphicx}
\usepackage{color} 

\begin{document}
\newcommand{\comment}[1]{}

\title{Precise in-situ tuning of a superconducting nanowire's critical current using high bias voltage pulses}

\author{Thomas Aref}
\author{Alexey Bezryadin}
\email{bezryadi@illinois.edu}
\affiliation{
University of Illinois at Urbana-Champaign
1110 W. Green St.
Urbana, IL 61801
}

\date{\today}

\begin{abstract}
We present a method for in-situ tuning the critical current (or switching current) and critical temperature of a superconducting nanowire using high bias voltage pulses. Our main finding is that as the pulse voltage is increased, the nanowires demonstrate a reduction, a minimum and then an enhancement of the switching current and critical temperature. Using controlled pulsing, the switching current of a superconducting nanowire can be set exactly to a desired value. These results correlate with in-situ transmission electron microscope imaging where an initially amorphous nanowire transforms into a single crystal nanowire by high bias voltage pulses. 
\end{abstract}


\maketitle


Superconducting nanowires have been proposed as candidates for various solid state qubit implementations \cite{mooij2005, mooij2006, ku2010}. We describe a post-fabrication technique using high bias voltage pulses that allows in-situ precise control over the critical current which may be highly instrumental in developing superconducting nanowire qubits.

\begin{figure}
\includegraphics[width=3.2 in]{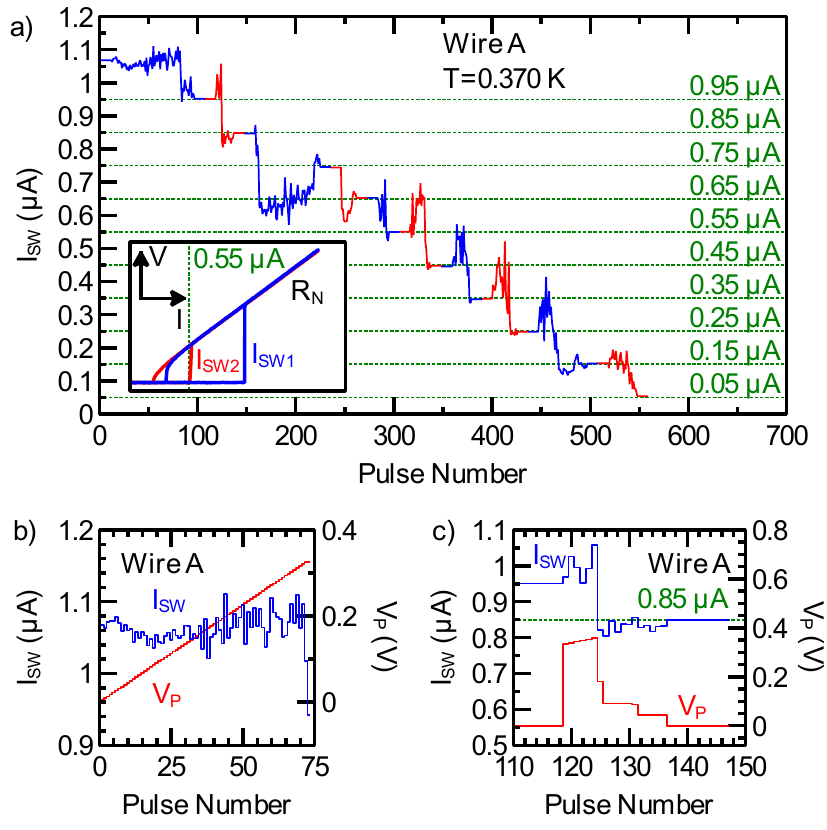}
\caption{a) $I_{SW}$ can be set exactly using a combination of large and small pulses. 
The flat regions correspond to the set value of $I_{SW}$ where no pulsing is applied. The noisy regions corresponds to $I_{SW}$ being set to the desired value as pulsing is applied.  In this example, starting $I_{SW}$ was $1.07 \mu$A and the chosen target values are shown by the green dotted lines. The inset shows how $I_{SW}$ and $R_N$ are extracted from the VI curves. $I_{SW1}=1.07 \mu$A at pulse number 0. $I_{SW2}= 0.55 \mu$A at pulse number 306. b) Close up on effect of pulsing on $I_{SW}$ as $V_P$ is increased from $0$ to $0.326$V. As $V_P$ grows, $I_{SW}$ becomes increasingly stochastic until at large enough pulses we observe a strong downward trend. In this case, the drop in $I_{SW}$ occurs at pulse number $73$ for $V_P=0.326$V.  c) Setting of $I_{SW}$ to 0.85$\mu$A with accompanying voltage pulses. At first large pulses are applied to get $I_{SW}$ near the desired value. Then smaller pulses are applied to bounce the $I_{SW}$ to the exact value desired.
}
\label{fig:stochasticIsw}
\end{figure}

As high bias pulses are applied to superconducting MoGe nanowires, we observe a decrease of switching current, $I_{SW}$, which is measured at a fixed temperature, typically 0.3K. Analysis of the rate of Little's phase slips indicates that pulsing also causes a decrease in the critical temperature, $T_C$, of the wire.
 An interesting application of this effect is to precisely control the switching current of a superconducting nanowire in-situ.  We demonstrate that the switching current can be set to within approximately 10 nA of a desired value (see figure \ref{fig:stochasticIsw}a). The switching current cannot be more accurately defined as it is natural stochastic due to thermal and quantum fluctuations \cite{sahu2009}. 

The nanowires were fabricated using the method of molecular templating \cite{bezryadin2000}. Both single-walled (SWNT) and multi-walled (MWNT) carbon nanotubes suspended across a trench were coated with Mo$_{76}$Ge$_{24}$ to form the nanowires (no difference was observed in behavior between SWNT and MWNT). Single voltage pulses were applied by using relays to switch from a quasi-four probe measurement to a single voltage pulse source. No effect on any nanowire was observed from just switching the relays without pulse application. The pulse length was kept fixed at 100 $\mu$s and was checked to transmit through the filtering of the cryostat to the sample.

From voltage vs current measurements, we can extract the switching current, $I_{SW}$, and the normal resistance, $R_N$, (see inset of figure \ref{fig:stochasticIsw}a) versus pulse number or pulse voltage, $V_P$, applied to the nanowire. The effect as $V_P$ is increased from $0$ to $0.326$V in $5$mV steps is shown in figure \ref{fig:stochasticIsw}b.   As $V_P$ is increased, we observe an increasing stochasticity of $I_{SW}$ which quickly becomes greater than the natural stochasticity of the switching current. As $V_P$ increases further, we see a sharp, downward trend of $I_{SW}$.  It should be noted that $I_{SW}$ was always measured sufficiently after the voltage pulse was finished that the wire had time to completely equilibrate to the bath temperature. Thus the observed changes in $I_{SW}$ are due to the voltage pulse permanently altering the wire and not heating effects of the high bias pulse. We can use this combined downward trend and increased stochasticity to precisely set $I_{SW}$ to a desired value. In figure \ref{fig:stochasticIsw}a, $I_{SW}$ is set to ten values chosen uniformly from $0.95 \mu$A to $0.05 \mu$A. An example of a pulse sequence used to set $I_{SW}$ is shown in figure \ref{fig:stochasticIsw}c. Large pulses are used to approach the desired value and then smaller pulses are used to `bounce' the switching current to within $\approx 10$nA of the desired value. For each of the ten chosen target values, the switching current was set to the desired value. 

\begin{figure}
\includegraphics[width=3.3 in]{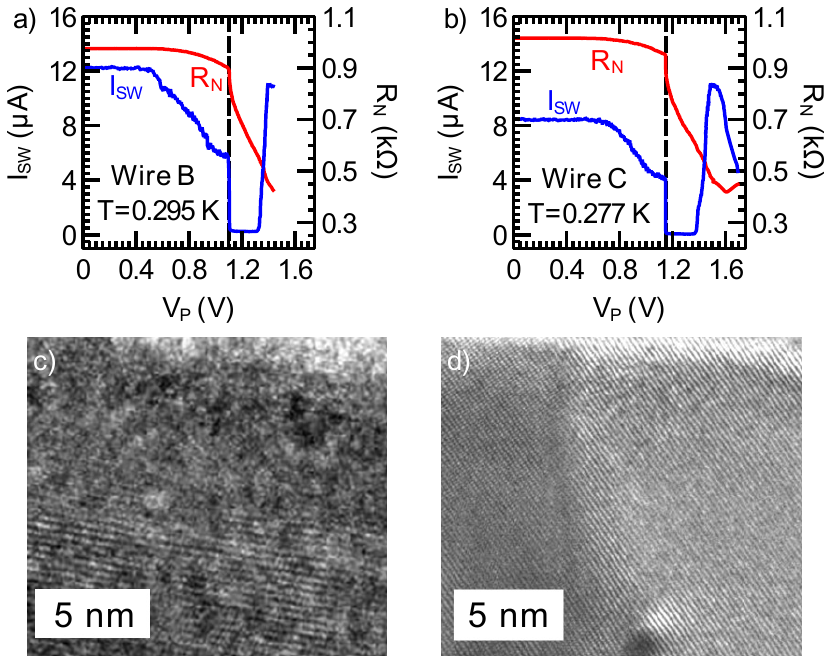}
\caption{Switching current, $I_{SW}$, and normal resistance, $R_N$, vs the maximum pulse voltage, $V_P$, applied to the wire. The blue line is $I_{SW}$ which decreases and then increases with increasing pulse amplitude. The red line is $R_N$ which stays roughly constant and then decreases. The dashed line indicates where both $R_N$ and $I_{SW}$ begin to rapidly decrease.  Applying pulses smaller than the maximum previous applied pulse does not lead to a significant change in $I_{SW}$.  a) A nanowire with starting $I_{SW}=12.2 \mu$A. The dashed line is at $1.105$V. $I_{SW}$ returns to $11.0 \mu$A before the wire abruptly breaks. b) A different nanowire with starting $I_{SW}=8.5 \mu$A. The dashed line is at $1.150$V. $I_{SW}$ returns to a maximum of $10.9 \mu$A (which is greater than the starting $I_{SW}$) before decreasing again until the wire breaks. c) In-situ TEM imaging of a nanowire (at room temperature) exposed to pulsing. Before pulsing, the nanowire is amorphous (the crystalline structure visible is the underlying MWNT supporting the metal of the wire). d) After pulsing to $3.735$V and at the point of breaking, the nanowire has been completely  crystallized.  
}
\label{fig:IswandRn}
\end{figure}

As $V_P$ is increased further,  $I_{SW}$ saturates at a minimum and then counterintuitively begins to increase, returning to values similar to or even exceeding the starting value of $I_{SW}$. 
 The drop, saturation and recovery of $I_{SW}$ can be see in figure \ref{fig:IswandRn}a and \ref{fig:IswandRn}b. The initial drop of $I_{SW}$ does not have a corresponding change in $R_N$. When $I_{SW}$ reaches a minimum and then begins to increase, $R_N$ begins to drop.  This behavior was reproduced on many nanowires of which figure \ref{fig:IswandRn} contains two examples.
SEM imaging of wires before and after pulsing do not indicate any formation of weak links due to pulsing. 

Weak links would only account for the reduction of switching current and not explain the observed recovery of $I_{SW}$ with increased pulse voltage. In-situ TEM imaging of a nanowire during pulsing indicates the initially amorphous nanowire becomes crystallized as pulsing proceeds. In figure \ref{fig:IswandRn}c, an unpulsed, amorphous Mo$_{76}$Ge$_{24}$ wire is shown with the crystal structure of the underlying MWNT (measured line spacing $3.2 \pm 0.1$ \AA) visible in the bottom half of the image. The crystallization begins by forming polycrystals in the center of the nanowire and gradually annealing the nanowire into a single crystal (measured line spacing $2.2 \pm 0.1$ \AA) as shown in figure \ref{fig:IswandRn}d.

\begin{figure}
\includegraphics[width=3.3 in]{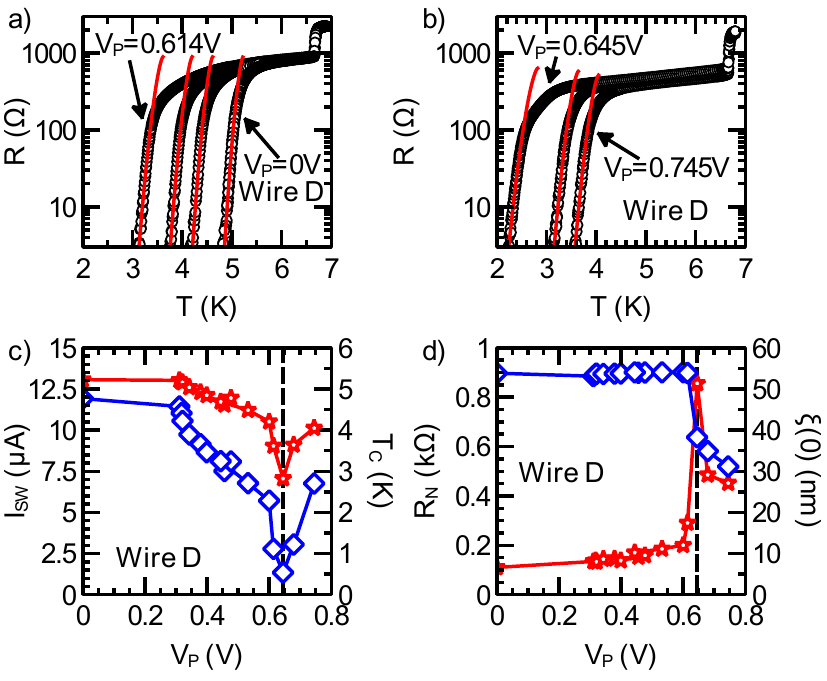}
\caption{Resistance vs temperature curves and fits. a) Four RT curves taken as pulsing generally drives $I_{SW}$ down. From right to left, the corresponding pulse voltages are $0$V, $0.456$V, $0.600$V and $0.614$V. The red curves are best fits to a TAPS model. b) Three RT curves taken when $I_{SW}$ returns for the same wire as shown in a.  From left to right the pulse voltages are $0.645$V, $0.679$V and $0.745$V.  
c) $I_{SW}$ and $T_C$ vs $V_P$ for the wire whose RT curves are shown in a and b. The diamonds correspond to $I_{SW}$ and the stars correspond to $T_C$. The dotted line is at $0.645$V where the turn around from decreasing to increasing behavior occurs. d) $R_N$ and $\xi(0)$ vs $V_P$. The diamonds correspond to $R_N$ and the stars correspond to $\xi(0)$.  Initially $R_N$ is flat while $\xi(0)$ shows a growing trend. After $V_P$ reaches $0.645$V, $R_N$ begins to drop and $\xi(0)$ shows a maximum and saturates to a value higher than the starting value. 
}
\label{fig:RTcurves}
\end{figure}

 The resistance versus temperature curves taken after a series of pulsing are shown in figure \ref{fig:RTcurves}a and \ref{fig:RTcurves}b. 
 The critical temperature, $T_C$, of the nanowire decreases as pulse voltage increases saturating at a minimum. $T_C$ is defined as a fitting parameter in the best Little fit (discussed in detail below). Further increase of pulse voltage results in the increase of $T_C$ and a drop in $R_N$ (see figure \ref{fig:RTcurves}c and \ref{fig:RTcurves}d). 

Most forms of crystalline MoGe have lower $T_C$'s than amorphous MoGe \cite{ghosh1977} so it is not surprising that the crystallization of MoGe would reduce the wire's critical temperature thereby reducing critical current at a given temperature.  It can be expected that any crystallization or segregation of the MoGe alloy from the large current pulse would produce a reduction of $T_C$. 

 The observed crystallization of the MoGe is caused by a combination of electromigration\cite{park1999, strachan2008, heersche2007} and Joule heating induced thermal effects. It appears that thermal effects are dominant since we observe the appearance of crystals at the center hottest spot of the wire and also do not observe the weak link formation associated with electromigration.  As a rough approximation of the temperature of the nanowire, we can write the applied voltage as a function of temperature (assuming a constant resistivity): $V^2/4=L(T^2-T_0^2)$
where $V$ is the voltage of the pulse, $L=2.4 \times 10^{-8} W \Omega/K^2$ is the Lorenz number, $T$ is the temperature of the wire center and $T_0$ is the temperature of the electrodes \cite{holm1967}. Typical values (V=0.5V) gives us an estimated temperature of $T= 1725$K close to the crystalizing temperature of MoGe.

The particular crystal form of MoGe closest to our starting concentration of Mo$_{76}$Ge$_{24}$ is Mo$_3$Ge which is an A15 compound (known to have high $T_C$'s). Studies on Mo$_3$Ge reveal that its $T_C$ is highly dependent on formation conditions (i.e. it can have a very low $T_C$) but under the correct formation conditions, the $T_C$ can exceed 5.7K (comparable to the critical temperature of the starting amorphous MoGe) \cite{ghosh1977}.  
It has been observed experimentally that Mo$_3$Ge can be generated by heating amorphous MoGe to high temperatures \cite{searcy1952}. Thus we propose that the return and sometimes higher $I_{SW}$ is caused by the formation of relatively well-ordered crystal Mo$_3$Ge from Joule heating by the high bias pulses. Comparison of TEM images measured line spacing (see figure \ref{fig:IswandRn}c and d) to x-ray diffraction data \cite{searcy1952} conclusively indicate that the crystal formed is indeed Mo$_3$Ge.
To compare to previous experiments on nanowires we use a phenomenological Little thermally activated phase slip (TAPS) model \cite{bezryadin2008}:
$R(T)=R_N \exp\left ( -\Delta F(T)/k_BT \right )$ where  $R_N$ is the normal resistance of the nanowire, $\Delta F(T)$ is the free energy barrier for phase slips, $k_B$ is the Boltzmann constant and $T$ is temperature \cite{chu2004, tinkham1996, langer1967, bezryadin2008}. The temperature dependence of the free energy barrier is accurately given at all temperatures by the Bardeen formula  \cite{bardeen1962}: $\Delta F(T)=\Delta F(0)  [1-(T/T_C )^2)^{3/2} \label{eq:bardeen}$
where we can express $\Delta F(0)$ using experimentally accessible parameters \cite{tinkham2002}. This model is used to produce the fits shown in figure \ref{fig:RTcurves} where the fitting parameters are critical temperature, $T_C$, and coherence length, $\xi(0)$. 

 As shown in figure \ref{fig:RTcurves}c, the decrease and return of $I_{SW}$ corresponds to a drop and return of $T_C$ as expected. In figure \ref{fig:RTcurves}d, we see that $R_N$ is stable and coherence length is gradually increasing as would be expected from the corresponding decrease in $T_C$ \cite{tinkham1996}. When $I_{SW}$ saturates at a minimum and begins to increase,  $R_N$ starts to decrease. Likewise, the coherence length returns from a maximal value (the maximum being due to a highly reduced $T_C$) and approaches a value higher than its initial value. This is reasonable since as the wire becomes well-ordered crystalline Mo$_3$Ge, we anticipate an enhanced coherence length due to the longer mean free path of the crystal compared to amorphous MoGe. From a simple Drude model of resistivity, a longer mean free path also implies a decreased normal resistance in agreement with the observed drop in resistance.

In conclusion, we demonstrate that controlled high bias pulsing can be used to precisely set the switching current of the nanowire and that the counterintuitive decrease and increase of the switching current with increasing pulse voltage is well explained by crystallization induced by Joule heating.

\begin{acknowledgments}
We thank Jian-Guo Wen for help with the TEM analysis. This material is based upon work supported by NSF-DMR 10-05645 and by the U.S. Department of Energy under grants DE-FG02-07ER46453 and DE-FG02-07ER46471 through the Frederick Seitz Materials Research Laboratory at the University of Illinois at Urbana-Champaign.
\end{acknowledgments}

\end{document}


\newcommand{\comment}[1]{}

\title{Precise in-situ tuning of a superconducting nanowire's critical current using high bias voltage pulses: Supplementary Information}

\author{Thomas Aref}
\author{Alexey Bezryadin}
\email{bezryadi@illinois.edu}
\affiliation{
University of Illinois at Urbana-Champaign
1110 W. Green St.
Urbana, IL 61801
}

\date{\today}

\begin{abstract}
This document contains materials and methods, measurement details, additional figures and explanation of the analysis.
\end{abstract}


\section{Supplementary Information}
Standard samples to be measured at cryogenic temperatures were prepared using molecular templating\cite{bezryadin2000}. Fluorinated single wall carbon nanotubes (SWNTs) are suspended across a trench in a Si substrate coated with SiO$_2$ and SiN films. Mo$_{76}$Ge$_{24}$ is deposited by DC sputtering forming a nanowire by using the nanotube as a nanoscaffold. Pattern definition by photolithography and the undercut of the trench allow only one conductance path, the nanowire, to be formed. 

\begin{figure}
\includegraphics[width=3.3 in]{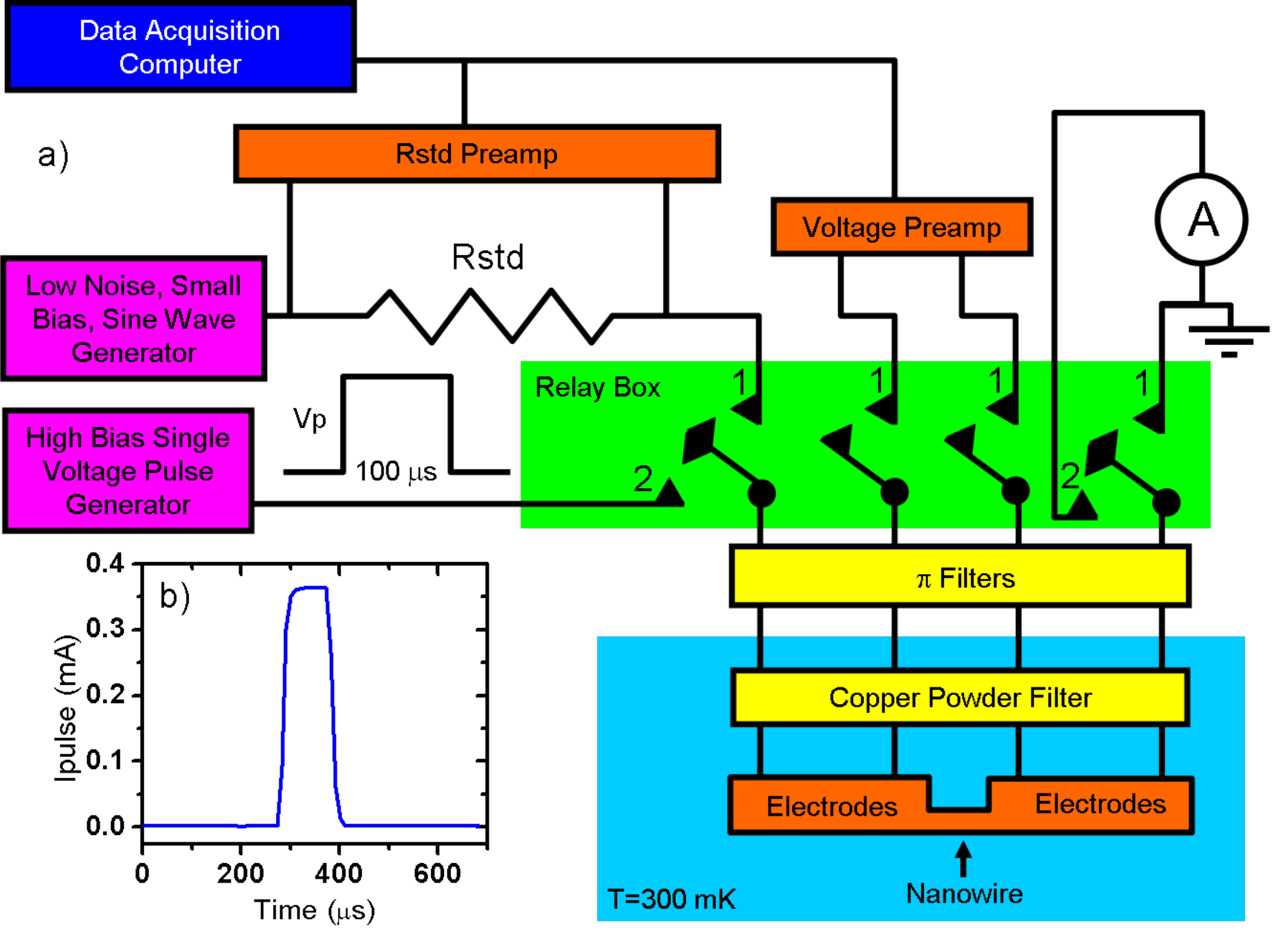}
\caption{Experiment Set-up: a) 4 relays (voltage controlled switches) are used to switch between measurement and pulsing mode. In measurement mode (all relays in position 1), we can measure either voltage vs current curves (amplitude of current $\approx $1-10$\,\mu$A) or resistance vs temperature curves (amplitude of current $\approx $10-20$ \,$nA). A sine wave generator connected through $R_{std}=$ 0.1-1$\,$M$\Omega$ forms a current source connected to the left lead. A small voltage ($\approx$1-10$\,$mV) is measured on the two center leads. The right current lead is grounded. In pulsing mode (relays in position 2), a single high bias voltage pulse ($\approx$0.1-1$\,$V) is sent in on the left current lead, the two center leads are disconnected and the pulse can be detected on the right lead using an ammeter ($\approx$0.1-1$\,$mA). b) Example measured current going through nanowire from a high bias voltage pulse measured with the ammeter. The pulse is 100 $\mu$s long and there are minor amounts of rounding of the pulse due to filtering in the cryostat.
}
\label{fig:setup}
\end{figure}

 The superconducting properties of the nanowires were measured in a He$-4$ (base temperature 1.5 K) or He$-3$ system (base temperature 0.3 K). The nanowires were measured in a standard current biased set-up with a low noise voltage source feeding a large value standard resistor $R_{std}$ serving as a current source and separate voltage probes. The four-probe measurement is of the superconducting electrodes and not the nanowire itself but the superconducting electrodes are seamlessly connected to the nanowire so we label it a quasi-four probe measurement. In order to protect sensitive measurement equipment from high bias pulses (1 V or more) and to allow application of a voltage bias rather than a current bias pulse, a switching system was employed to switch between measurement mode and pulsing mode (see Figure \ref{fig:setup}a). The wire was pulsed between sensitive measurements in order to change its morphology.

Both manually operated switches and automated relays (voltage powered switches controlled by a computer) were used. No difference in behavior of the nanowires was observed between the two. The relays were low bias, latching relays powered by a Keithley electrometer controlled by the measurement computer through GPIB. The latching design of the relays allows the power to the relays to be removed without affecting the switch position of the relays. To test the relays, repeated switches were made with no pulse application. No effect on any nanowire was observed from just switching back and forth without pulse application. Square pulses were applied using a data acquisition (DAQ) card. Pulse duration was kept at 100 $\mu$s and pulse voltage amplitude was varied. Pulses of this length transmit fairly well through the filtering system on the cryogenic measurement systems maintaining their square shape with minimal rounding (see Fig. \ref{fig:setup}b).
 
 We have not systematically explored the effect of different length pulses (or different shaped pulses) but we do not expect significant dependence on these two factors for the following reasons. The response time of the nanowire should be on the order of picoseconds (the capacitance of the electrodes is on the order of a few fF \cite{bollinger2006} while the resistance is approximately 1 k$\Omega$ giving a RC time constant of approximately 1-10 picoseconds) so the wire will have reached equilibrium current early in the pulse. The wire is expected to reach its maximum temperature (due to Joule heating) during the pulse and cool back to base temperature after the pulse within 10-100 ns \cite{sahu2009} so it should be well thermally equilibrated early in the pulse as well. It should be noted that our relay switching speed (approximately 1 second) is not fast enough to allow us to capture the cooldown back to base temperature after the pulse and the wire is well thermalized before switching currents are measured after a pulse.

\begin{figure}
\includegraphics[width=3.3 in]{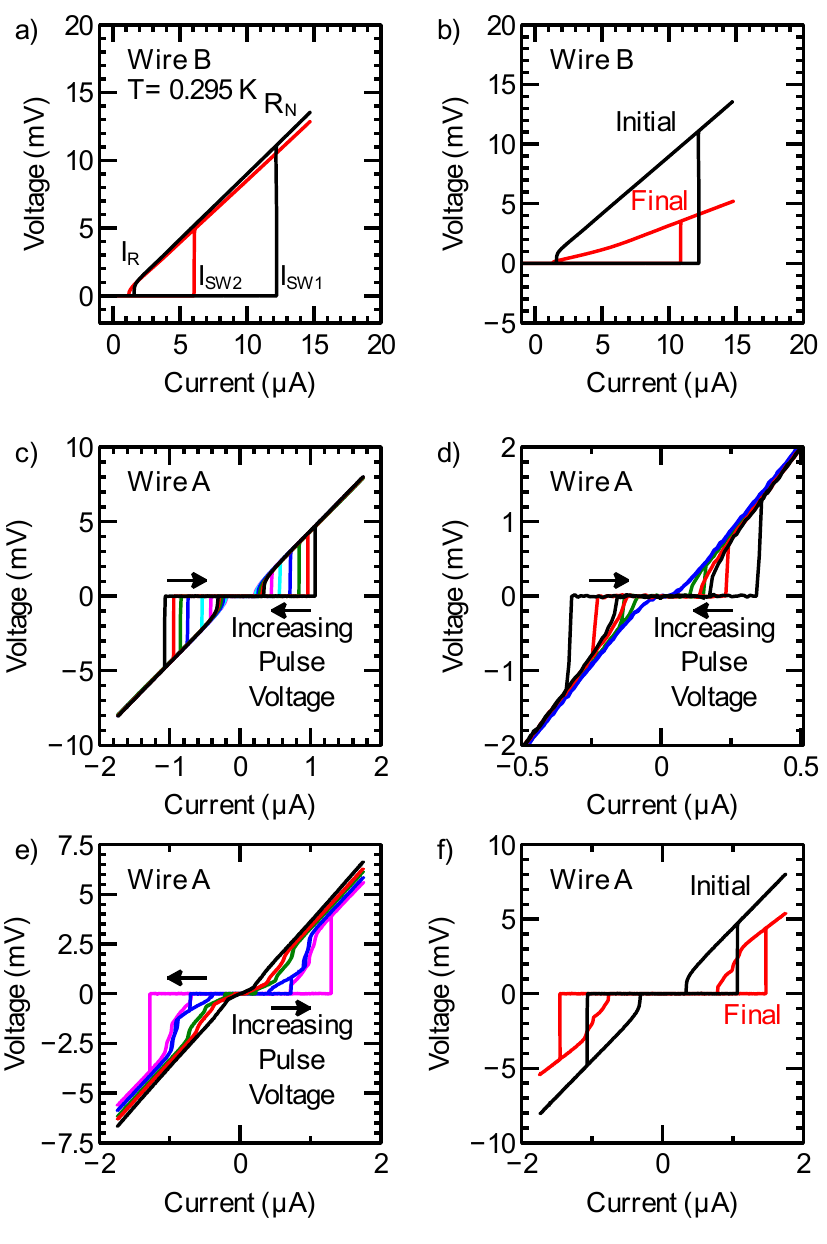}
\caption{Voltage vs current (VI) curves demonstrating effect of high bias pulses. a) Initial application of high bias pulses decreases the switching current from $I_{SW1}$ to $I_{SW2}$ while minimally changing $R_N$ and $I_R$. b) Further pulsing results in the return of $I_{SW}$ and a drop in $R_N$. `Initial' is the same curve as the one shown in black in graph a. `Final' is the last VI curve before the sample broke. c) A different nanowire with a smaller initial $I_{SW}$. This graph shows many VI curves to indicate the gradual decrease of the $I_{SW}$ as increasing pulses are applied d) Application of higher pulses results in a loss of hysteresis of the VI curve. However, the VI curve retains a flat, superconducting region with apparently non-zero critical current e) Still higher pulses results in the return of $I_{SW}$ with a drop in $R_N$ and evidence of phase slip centers. f) Still higher pulsing produces an almost uniform VI curve with $I_{SW}$ exceeding the original switching current and a further drop in $R_N$. `Initial' is the same curve as the one shown in black in graph c. `Final' is the last VI curve before the sample broke.
}
\label{fig:VIcurves}
\end{figure}

In measurement mode, a low bias sine wave signal current source is applied to the nanowire and voltage is measured separately using the quasi-four probe measurement described previously (see figure \ref{fig:setup}a). Typical voltage versus current (VI) curves and the effects of pulsing on them are shown in figure \ref{fig:VIcurves}. $I_{SW}$ initially decreases with minimal change in $R_N$ and $I_R$  and the VI curves maintain single hysteretic loops characteristic of homogeneous wires.  The hysteresis in the VI curve disappears as the switching current goes to a minimum (see figure \ref{fig:VIcurves}d). A flat, superconducting region indicates a non-zero critical current always remains. Higher pulsing results in a return of the hysteretic VI curve with now increasing $I_{SW}$ and decreasing $R_N$ as pulse voltage is increased. When $I_{SW}$ returns, we generally observe phase slip centers in the VI curves (see figure \ref{fig:VIcurves}e) indicating less homogeneous nanowires. As pulse voltage is further increased, these phase slip centers gradually disappear. The wire can return to a $I_{SW}$ approaching the starting $I_{SW}$  (see figure \ref{fig:VIcurves}b) or even exceeding it  (see figure \ref{fig:VIcurves}f) and as shown in the paper. Graphing multiple VI curves on top of each other quickly becomes a tedious and confusing way to present the data so $I_{SW}$ and $R_N$ versus $V_P$ was the preferred representation of the data in the paper.  It should be noted that as the switching current goes through its minimum it's exact value is not as exactly determined by the threshold detection scheme as for curves with higher switching current values. In order to minimize the natural stochasticity of the switching current \cite{sahu2009}, we averaged over $100$ switching current measurements between pulses.
 
For resistance vs temperature (RT) curves, the low bias current signal was reduced from $\approx$1-10$\,\mu$A to $10-20\,$nA to measure the RT curve in the linear regime. The RT curves generally demonstrates one transition indicative of a homogeneous wire with fitting parameters similar to unpulsed nanowires . 

\begin{figure}
\includegraphics[width=3.3 in]{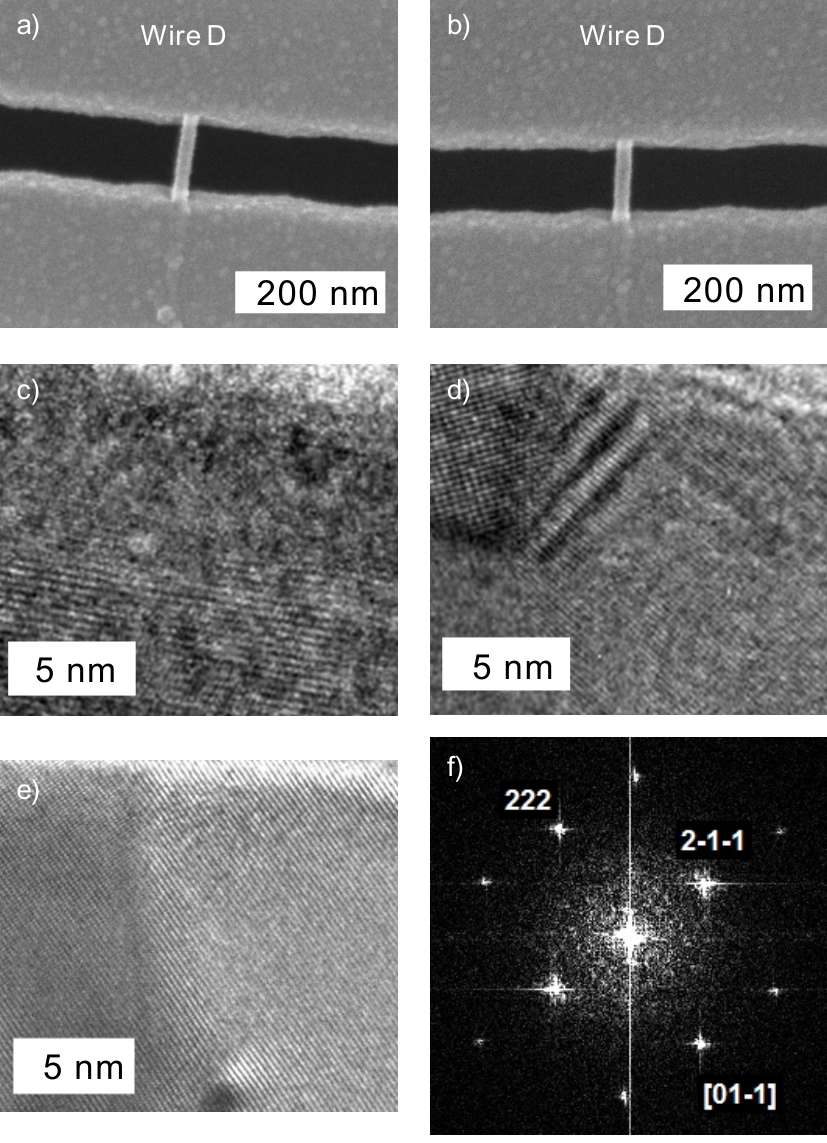}
\caption{SEM and TEM analysis a) SEM image of a nanowire before pulsing at cryogenic temperatures b) the same nanowire as a after pulsing at cryogenic temperatures with a maximum pulse voltage applied of $588$mV. Switching current of the wire was reduced from $10.3\mu$A down to a minimum and then back to $8.5\mu$A. Note no obvious weak links are visible. c) In-situ TEM imaging of a nanowire (at room temperature) exposed to pulsing. Before pulsing, the nanowire is amorphous (the crystalline structure visible is the underlying MWNT). d) After pulsing to $2.935$V, the wire becomes polycrystalline. e) After pulsing to $3.735$V, the nanowire becomes a single crystal. f) Diffraction pattern of the crystal shown in e. 
}
\label{fig:TEMpics}
\end{figure}

SEM imaging before and after pulsing show virtually no change in the nanowire ruling out the formation of obvious weak links (small scale weak links below the resolution of the SEM are still possible). In figure \ref{fig:TEMpics}a and b a wire pulsed down to a minimum switching current and then back to high switching current is shown. There are only small differences between the images. Similar experiments on a wire pulsed to a minimum switching current again did not reveal any weak link. 

To more thoroughly study the pulses effect on the nanowires, we turn to in-situ TEM experiments. TEM experiments require different samples than those described previously. Most importantly, the nanowire must be across an open slit for TEM observation. We deposit multi-walled carbon nanotubes (MWNTs) across TEM compatible slits to generate these samples \cite{aref2009}. We use a KOH etch to fabricate a V-shaped cut in a silicon chip coated on both sides with 100 nm of silicon nitride. The V-shaped cut almost pierces the chip except for approximately 5 microns of remaining silicon. This silicon is cracked by sonicating in deionized water for less than a second. A 30-60 second KOH etch removes the cracked silicon leaving an approximately 100 nm wide silicon membrane. This membrane is removed by RIE etching from the etch pit side. The membrane is supported during the RIE step by a piece of polydimethylsiloxane (PDMS). In the method previously described \cite{aref2009} we removed the silicon nitride entirely and oxidized the silicon to form an insulating layer. By etching the silicon nitride from the etch pit side, we are able to use the silicon nitride as the insulating layer thus skipping the oxidizing step.

 With these samples, we were able to perform in-situ TEM experiments to directly determine the high bias voltage pulses effects on metal coated nanotubes. MWNTs were used rather than the SWNTs used for regular nanowire samples. MWNTs are more robust and rigid and thus can more easily be deposited on the TEM compatible slits \cite{aref2009}. We checked that the change in scaffold does not affect the pulsing behavior at cryogenic temperatures. The in-situ TEM experiments must be done at room temperature while superconducting measurements must be done at cryogenic temperatures. This change in base temperature can be safely neglected as both experiments are performed under vacuum and the nanowire itself is expected to reach a high temperature ($\approx 2000$K) under high bias voltage ($\approx 0.5\,$V).
 
From the in-situ TEM experiments we see that initially the wire develops a polycrystalline section which expands as increasing pulses are applied. The crystals do not necessarily remain static for the duration of the experiment but rather are dynamic entities that develop and change. The polycrystalline nature of the wire gradually becomes dominated by fewer and fewer crystal domains and becomes an almost perfect single crystal nanowire. It should be noted that although inhomogeneities such as grain boundaries appear in the wire, the overall diameter of the wire does not appear to be significantly altered. The crystallization of MoGe from a high voltage pulse is not surprising in light of similar crystallization obtained by exposure of MoGe nanowires to electron beam radiation \cite{remeika2005}. To avoid electron beam induced crystallization in our TEM images, dosage from the electron beam was minimized for all TEM images. As further evidence that the high bias pulses and not the electron beam of the TEM were responsible for the crystallization observed, the same crystallization was seen in nanowires constantly imaged during the pulse process as was seen in wires that were not imaged until the pulse process was complete. In the first TEM image, the multiwalled nanotube (with wall spacing $3.3$ \AA) covered with amorphous Mo$_{76}$Ge$_{24}$ is visible (see figure \ref{fig:TEMpics} c). The measured line spacing in the image is $3.2 \pm 0.1$ \AA. After some pulsing, a polycrystalline structure is visible with the predominant line spacing being $2.2 \pm 0.1$ \AA. Only in the upper left hand corner is the line spacing different $2.5 \pm 0.1$ \AA (see figure \ref{fig:TEMpics} d). In the final TEM picture the single crystal line spacing is $2.2 \pm 0.1$ \AA (see figure \ref{fig:TEMpics} e).  TEM imaging shows that a polycrystalline morphology appears with voltage pulsing.
 Following the work of Rogachev et al. \cite{rogachev2003}, we can expect these polycrystalline wires to maintain homogeneous wire behavior and can fit them using standard nanowire theory. Also in agreement with these previous results, we see phase slip centers develop in the VI curve (see figure \ref{fig:VIcurves}e) at temperatures near $T_C$ (as we are changing $T_C$ while keeping $T$ fixed, these are most evident when $T_C$ is small). 



By comparing x-ray diffraction data\cite{searcy1952} for Mo$_3$Ge  and our TEM images, we can confirm that Mo$_3$Ge is being formed by the pulses. In agreement with the x-ray diffraction data, our most commonly observed orientations are 210 (2.1993 \AA from x-ray diffraction) and 211 (2.0031 \AA from x-ray diffraction). Our measurement of 2.2 \AA{} by itself is not accurate enough to tell the difference between these two orientations. From further analysis of the TEM images, we retrieve that the spacing of the crystal in the 222 direction is 1.5 \AA{} (compared to 1.4215 \AA{} from x-ray diffraction). For the 211 direction, we measure 2.1 \AA{} (compared to 2.0031 \AA{}  from x-ray diffraction). On the upper left hand corner of the polycrystal, we observe a spacing of 2.5 \AA{} (compared to 2.4557 \AA{} from x-ray diffraction) for the 200 and 3.3 \AA{} for the 110 direction (compared to 3.4724 \AA{} from x-ray diffraction). Both these orientations are significantly less common than the others which dominate the images.  In all cases, there is significant correlation between our values and the x-ray diffraction data indicating that the crystal we are generating is indeed Mo$_3$Ge \cite{searcy1952}.


The coherence length of a dirty superconductor's dependence on mean free path and critical temperature is given by: $ \xi(0) = 0.855 [(l \hbar v_F)/(1.76 \pi k_B T_c)]^{1/2} \label{eq:xitheory}$  where $l$ is the mean free path of MoGe ($\approx$ 3.5 \AA) and $v_F$ is the Fermi velocity ($\approx 1 \times 10^6$ m/s)\cite{tinkham1996}. This equation is the source of the qualitative analysis of changes in coherence length given in the paper

The expression relating the free energy barrier to experimentally determined parameters is: $ \Delta F (0)=A_T \frac{R_q}{R_N}\frac{L}{\xi(0)}k_B T_C \label{eq:DeltaFzero}$
 where $A_T=1.76 \sqrt{2}/3=0.83$, $R_Q$ is the resistance quantum, $R_N$ is the normal resistance of the nanowire (which we define as the resistance immediately after the film goes superconducting), $L$ is the length of the nanowire (which can be determined from SEM imaging), $T_C$ is the critical temperature and  $\xi(0)$ is the coherence length\cite{tinkham2002}. We use $T_C$ and $\xi(0)$ as fitting parameters. We have checked that there is a high degree of overlap between fitting parameters for pulsed nanowires and unpulsed nanowires indicating that it is fair to treat a pulsed nanowire using a similar theoretical description to an unpulsed nanowire.

%
